\documentclass[twocolumn,showpacs,amsmath,amssymb]{revtex4}
\usepackage[dvips]{graphics}
\def\imo{i}

\begin{document}
\title{Massive charged scalar field in the Kerr-Newman background II: Hawking radiation}
\author{R. A. Konoplya}\email{konoplya_roma@yahoo.com}
\affiliation{DAMTP, Centre for Mathematical Sciences, University of Cambridge, Wilberforce Road, Cambridge CB3 0WA, United Kingdom.}
\author{A. Zhidenko}\email{zhidenko@physik.uni-frankfurt.de}
\affiliation{Institut f\"ur Theoretische Physik, Johann Wolfgang Goethe-Universit\"at, Max-von-Laue-Str. 1, 60438 Frankfurt, Germany}
\affiliation{Centro de Matem\'atica, Computa\c{c}\~ao e Cogni\c{c}\~ao, Universidade Federal do ABC (UFABC), Rua Aboli\c{c}\~ao, CEP: 09210-180, Santo Andr\'e, SP, Brazil}

\begin{abstract}
  We perform accurate calculations of the energy-, momentum-, and charge-emission rates of a charged scalar field in the background of the Kerr-Newman black hole at the range of parameters for which the effect is not negligibly small and, at the same time, the semiclassical regime is, at least marginally, valid. For black holes with charge below or not much higher than the charge accretion limit $Q \sim \mu M/e$ (where $e$ and $\mu$ are the electron's mass and charge), the time between the consequent emitting of two charged particles is very large. For primordial black holes the transition between the increasing and decreasing of the ratio $Q/M$ occurs around the charge accretion limit. The rotation increases the intensity of radiation up to three orders, while the effect of the field's mass strongly suppresses the radiation.
\end{abstract}
\maketitle

\section{Introduction}
The intensity of the Hawking radiation of charged and neutral particles from a black hole has been usually estimated within the semiclassical approximation. For a given mass of the black hole $M$, the particle of mass $\mu$ can be described still within the semiclassical approximation if the Compton wavelength $\lambda_c = 2 \pi /\mu$ is much smaller the black hole size $2 r_h \approx 4 M$ ($\hbar = c = G =1$):
\begin{equation}
\lambda_c \ll 2 r_h,\quad\Longrightarrow\quad\mu M \gg 1.
\end{equation}
When $\mu M \ll 1$ corrections due to essentially quantum nature of the field's propagation in the vicinity of the black hole must be taken into consideration \cite{Gibbons:1975kk}. Let us estimate physically relevant range of parameters, taking into account validity of the semiclassical approximation and realistic values of the black hole parameters.

The ration of the electron charge to its mass $e/\mu \sim 10^{21}$, while the the ratio of the black hole charge to its mass $Q/M$ must be tiny: a black hole cannot hold a large electric charge and once $e Q > \mu M$ no further freely falling charge can accrete onto the black hole. At the same time, Hawking radiation of solar mass  black holes is negligible, while hypothetic mini-black holes owing to possible extra dimensions must have such a small mass $M$, that no Standard Model particles can be described within $M \mu \gg 1$ regime. Thus, only for primordial black holes is the Hawking radiation both significant and can be described within the semiclassical approximation. In the SI units $\mu M = \mu M_{BH} G/c \hbar=\mu M_{BH}/m_{P}^2$ ($m_P\approx2\cdot10^{-8}kg$), so that for an electron ($\mu \sim 10^{-30} kg$), $\mu M \sim 1$ corresponds to $M_{BH} \sim 10^{15}$ kg, which is the regime of primordial black holes, including the black holes which could be candidates for the dark matter $M_{BH} > 10^{14} kg.$ \cite{DarkmatterBH}.

As the Planck charge is about $11$ times greater than the electron charge, even for the weakly charged black hole, the coupling of charges $e Q$ does not need to be small, and $e Q \sim 1$ corresponds to an electron in the vicinity of a black hole charged with $\sim 10^2$ electrons. Once the electric field on the black hole surface is sufficiently strong, an addition channel of radiation opens. If the work made by the electric field is larger than the energy necessary to transform a virtual pair of oppositely charged particles into the real one,
\begin{equation}
\lambda_c e E \geq 2 \mu, \quad E \sim Q/(2 M)^2, \quad \lambda_c= 2 \pi/\mu,
\end{equation}
or, simpler, if
\begin{equation}
e Q \gtrsim (\mu M)^2,
\end{equation}
then the Schwinger process of particle creation begins in the vicinity of a black hole.
Therefore, we shall consider the following range of parameters:
\begin{enumerate}
  \item $\mu M \geqslant 1$: semiclassical regime near its margin, which also includes primordial black holes for the Standard Model particles.
  \item $Q/M$ is very small, which means that, practically, one can take $Q=0$ everywhere in the wave equation, except for the coupling $e Q$.
  \item As $e/\mu \sim 10^{21}$ for an electron, this means that $e Q$ can range from $\sim 10^{-2}$ (one electron on the black hole surface) up to $(\mu M)$.
\end{enumerate}
Later we shall show that at the above range of parameters, the black hole should discharge from a given state $e Q > \mu M$ in a ``fragmentary'' way, so that a large period of time elapses between emitting of two consequent electrons making the state with $e Q > (\mu M)^2$ to be unlikely.
Thus, we can write down the physically relevant range of parameters as
\begin{equation}
10^{-2}< e Q < (\mu M)^2 \gtrsim 1, \quad Q/M \rightarrow 0.
\end{equation}

Although the effect of the black hole's discharge owing to the Hawking radiation of charged particles was considered a long time ago \cite{Gibbons:1975kk,Carter}, estimations of the emission rates were limited by the nonrotating case and either by $\mu M \ll 1$ or $\mu M \gg 1$ regimes \cite{Gibbons:1975kk,Page:1977um}.
The intensity of the Hawking radiation for a massive charged scalar field in the background of the Kerr-Newman black hole (immersed in a magnetic field) was considered in a recent paper by K. Kokkotas and us \cite{Kokkotas:2010zd}. There it was shown that the Zeeman shift of the particle energy in the magnetic field and the Faraday induction influence the intensity of emitting of charged particles. Nevertheless, the analysis in \cite{Kokkotas:2010zd} was done also for small $\mu M$ and with a number of approximations due to the inseparability of variables in the whole space at a nonzero magnetic field.
The emission of charged fields from higher-dimensional fields were considered also in \cite{Sampaio}.

In the present work, we shall complement the existing estimates of the intensity of Hawking radiation of massive charged fields and perform accurate calculations of energy-, momentum-, and charge-emission rates for a massive charged scalar field in the vicinity of a Kerr-Newman black hole. We shall show that for black holes with a charge below or not much higher than the charge accretion limit $Q \sim \mu M/e$, the time between consequent emitting of two charged particles is very large. The regime $M \mu \sim e Q$ is important for primordial black holes with a mass of about $10^{15} kg$, because the transition between the increasing and decreasing of the $Q/M$ ratio occurs in this regime. We shall show that the rotation increases the intensities of Hawking radiation up to three orders, while the effect of the field's mass strongly suppresses it.

The paper is organized as follows: Sec. II gives the basic formulas for the Kerr-Newman metric and a massive charged scalar field equation in its background. In Sec. III we briefly relate the numerical method for the finding of the grey-body factors. Sec. IV is devoted to an analysis of numerical data for the grey-body factors and various emission rates. Sec. V discusses the evolution of the black hole discharge owing to the Hawking radiation of charged particles. In the Conclusions we summarize the obtained results and mention some open questions.

\section{Kerr-Newman background}

In the Boyer-Lindquist coordinates the Kerr-Newman metric has the form
\begin{eqnarray}
ds^2 &=& \frac{\Delta_r}{ \rho^2}(dt-a\sin^2\theta d\varphi)^2-\rho^2
\left(\frac{dr^2}{\Delta_r}+\frac{d\theta^2}{\Delta_\theta}\right)\\\nonumber
&&-\frac{\Delta_\theta \sin^2\theta}{\rho^2}
[adt-(r^2+a^2)d\varphi]^2,
\end{eqnarray}
where
$$\Delta_r=(r^2+a^2)-2Mr+Q^2,$$
\begin{equation}
\rho^2=r^2+a^2\cos^2\theta,
\end{equation}
and $Q$ is the black hole charge; $M$ is its mass.
The electromagnetic background of the black hole is given by the
four-vector potential
\begin{equation}
A_{\mu}dx^{\mu}
   =-\frac{Qr}{\rho^2}(dt-a\sin^2\theta d\varphi).
\end{equation}
We shall parameterize the metric by the following three parameters: the event horizon $r_+$, the inner horizon $r_-$, and the rotation parameter $a$,
$$0\leq a^2/r_+\leq r_-\leq r_+.$$
The black hole's mass and charge are then
$$2M=r_++r_-,\qquad Q^2=r_+r_--a^2.$$

A massive charged scalar field obeys the equation
\begin{eqnarray}\label{KG}
&&\frac{1}{\sqrt{-g}}\frac{\partial}{\partial x^\alpha}\left(g^{\alpha\beta}\sqrt{-g}\frac{\partial\psi}{\partial x^\beta}\right)+2iqA_\alpha g^{\alpha\beta}\frac{\partial\psi}{\partial x^\beta}\\\nonumber&&+(\mu^2-q^2g^{\alpha\beta}A_\alpha A_\beta)\psi=0,
\end{eqnarray}
where $q$ and $\mu$ are the field's charge and mass, respectively.

One separates variables by the following ansatz:
\begin{equation}\label{anzats}
\psi = e^{-\imo \omega t + \imo m \phi} S(\theta) R(r)/\sqrt{r^2+a^2},
\end{equation}
where $S(\theta)$ satisfies the following equation:
\begin{eqnarray}\label{angularpart}
&&\left(\frac{\partial^2}{\partial \theta^2} + \cot \theta \frac{\partial}{\partial\theta} - \frac{m^2}{\sin^2 \theta} - a^2 \omega^2\sin^2 \theta\right.
\\\nonumber&&\left. + 2 m a \omega + \lambda - \mu^2 a^2 \cos^2 \theta \right) S(\theta) = 0,
\end{eqnarray}
and $\lambda$ is the separation constant.
The classical stability of a massive charged scalar field in the Kerr-Newman black hole and its characteristic quasinormal spectrum has been recently analyzed in \cite{Konoplya:2013rxa}.

This equation can be solved numerically for any fixed value of $\omega$ in the same way as the equation for a massive scalar field in the Kerr black hole background \cite{Konoplya:2006br}. When $\mu=0$, Eq.~(\ref{angularpart}) can be reduced to the standard equation for the spheroidal functions, and for any fixed value of $\omega$ the separation constant $\lambda$ can be found numerically, using the continued fraction method \cite{Suzuki:1998vy}. When the effective mass is not zero, the separation constant $\lambda(\omega, \mu)$ can be expressed, in terms of the eigenvalue for spheroidal functions $\lambda(\omega)$ \cite{Kokkotas:2010zd}, as
$$\lambda(\omega,\mu)=\lambda(\sqrt{\omega^2-\mu^2},0)+2ma(\sqrt{\omega^2-\mu^2}-\omega)+\mu^2a^2.$$
When $a=0$, one has $\lambda=\ell(\ell+1),~\ell=0,1,2\ldots$. For nonzero values of $a$, the separation constant can be enumerated by the integer multipole number $\ell\geq|m|$.

The radial function satisfies a Schr\"odinger-like equation,
\begin{equation}\label{radialpart}
\left(\frac{d^2}{d x^2} - V(x)\right) R(x) = 0,
\end{equation}
where $x$ is the tortoise coordinate,
$$dx=\frac{(r^2+a^2)}{\Delta_r}dr.$$
The effective potential has the form:
\begin{eqnarray}\nonumber
V&=&\frac{\Delta_r}{(r^2+a^2)^2}\left(\lambda + \mu^2 r^2+\frac{(r \Delta_r)'}{r^2 + a^2}-\frac{3 \Delta_r r^2}{(r^2 + a^2)^2}\right)
\\\label{Effective-potential}&&-\left(\omega - \frac{m a+eQr}{r^2 + a^2}\right)^2.
\end{eqnarray}

The asymptotics of the effective potential near the event horizon and at spatial infinity are
\begin{equation}
\begin{array}{lll}
V \rightarrow -\Omega^2, &~~r \rightarrow \infty~~(x\rightarrow \infty),&  \Omega = \sqrt{\omega^2 - \mu^2}>0,\\
V \rightarrow -\tilde\omega^2, &~~r \rightarrow r_+~~(x\rightarrow -\infty),& \displaystyle \tilde\omega = \omega - \frac{ma+eQr_+}{a^2+r_+^2}.
\end{array}
\end{equation}

\subsection{Reflection coefficients}

In order to calculate the emission rates of particles owing to Hawking radiation, one has first to solve the problem of classical scattering and  obtain the gray-body factors. This implies the posing of classical \emph{scattering boundary conditions}, that is, requiring of a purely ingoing wave at the event horizon ($x\rightarrow-\infty$),
\begin{equation}
R\propto \exp(-\imo\tilde\omega x) \propto (r-r_+)^{\displaystyle-\imo\tilde\omega/4\pi T_H},
\end{equation}
where $T_H$ is the Hawking temperature
\begin{equation}
T_H=\frac{\Delta'(r_+)}{4\pi(r_+^2+a^2)},
\end{equation}
and a linear combination of the ingoing and outgoing waves at spatial infinity ($r\rightarrow\infty$),
\begin{equation}
R\simeq Z_{in} \exp(-\imo\Omega x)+Z_{out} \exp(\imo\Omega x),
\end{equation}
where $Z_{in}$ and $Z_{out}$ are integration constants.

Note, that $\tilde\omega$ can be negative. This corresponds to the \emph{superradiant regime} \cite{Starobinsky} in which
\begin{equation}
0<\omega<\displaystyle\frac{ma+eQr_+}{a^2+r_+^2}.
\end{equation}

Let us introduce the new function, which is regular at the event horizon,
\begin{equation}
P(r)= R(r) \left(\frac{r-r_+}{r-r_-}\right)^{\displaystyle\imo\tilde\omega/4\pi T_H}.
\end{equation}
Then, choosing the integration constant as $P(r_+)=1$, we expand Eq.~(\ref{radialpart}) near the event horizon and find $P'(r_+)$, which completely fixes the initial conditions for the numerical integration. Then, we integrate Eq.~(\ref{radialpart}) numerically from the event horizon $r_+$ to some distant point $r_f\gg r_+$ and find a fit for the numerical solution far from the black hole in the following form:
\begin{equation}\label{fit}
P(r)=Z_{in} P_{in}(r)+Z_{out} P_{out}(r),
\end{equation}
where the asymptotic expansions for the corresponding functions are found by expanding (\ref{radialpart}) at large $r$ as
\begin{eqnarray}
P_{in}(r)&=&e^{-\imo\Omega r}r^{-\displaystyle\imo\sigma}\left(1+P_{in}^{(1)}r^{-1} + P_{in}^{(2)}r^{-2}+\ldots\right),\nonumber\\
P_{out}(r)&=&e^{\imo\Omega r}r^{\displaystyle\imo\sigma}\left(1+P_{out}^{(1)}r^{-1} + P_{out}^{(2)}r^{-2}+\ldots\right).\nonumber
\end{eqnarray}
The fitting procedure allows us to find the coefficients $Z_{in}$ and $Z_{out}$. In order to check the accuracy of the found coefficients, one should increase the internal precision of the integration procedure, the value of $r_f$, and the number of terms in the series expansion for $P_{in}(r)$ and $P_{out}(r)$, making sure that the values of $Z_{in}$ and $Z_{out}$ do not change within desired precision.

If the coefficients $Z_{in}$ and $Z_{out}$ are calculated, one can find the absorbtion probability
\begin{equation}\label{absorbtion}
|{\cal A}_{\ell,m} |^2=1-|Z_{out}/Z_{in}|^2.
\end{equation}
The above expressions will be used here for calculations of the energy-, momentum-, and charge-emission rates. This approach was used for calculations  of intensities of Hawking radiation for higher-dimensional simply rotating black holes \cite{Kanti:2009sn} and Gauss-Bonnet black holes \cite{Konoplya:2010vz}, showing an excellent agreement with the semianalytical approach \cite{Kanti-review} in the range of its validity.

\section{Scattering and Hawking radiation}

Here we suppose that a black hole evaporates adiabatically, so that the black hole can still be described by a stationary solution. We shall assume that the black hole is in thermal equilibrium with its surroundings in the sense of the canonical ensemble: the black hole temperature does not change between the emission of two consequent particles.

The energy-, angular momentum-, and charge-emission rates have the well-known form \cite{Hawking}

\begin{equation}\label{energy-emission}
-{{dM} \over {dt}} =
\sum_{\ell=0}^\infty \sum_{m=-\ell}^\ell \int{ \left| {\cal A}_{\ell, m} \right|^2 {\omega \over
{\exp (\tilde\omega /T_H ) - 1}}{{d\omega } \over {2\pi }}},
\end{equation}
\begin{equation}\label{charge-emission}
-{{dQ} \over {dt}} =
\sum_{\ell=0}^\infty \sum_{m=-\ell}^\ell \int{ \left| {\cal A}_{\ell, m} \right|^2 {e \over
{\exp (\tilde\omega /T_H ) - 1}}{{d\omega } \over {2\pi }}},
\end{equation}
\begin{equation}\label{momentum-emission}
-{{dJ} \over {dt}} =
\sum_{\ell=0}^\infty \sum_{m=-\ell}^\ell \int{ \left| {\cal A}_{\ell, m} \right|^2 {m \over
{\exp (\tilde\omega /T_H ) - 1}}{{d\omega } \over {2\pi }}}.
\end{equation}
\begin{figure*}
\resizebox{\linewidth}{!}{\includegraphics*{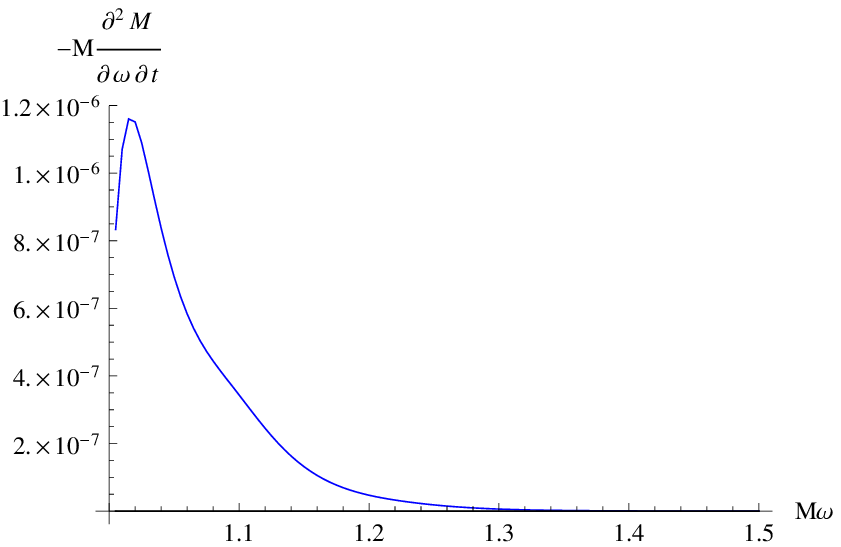}\includegraphics*{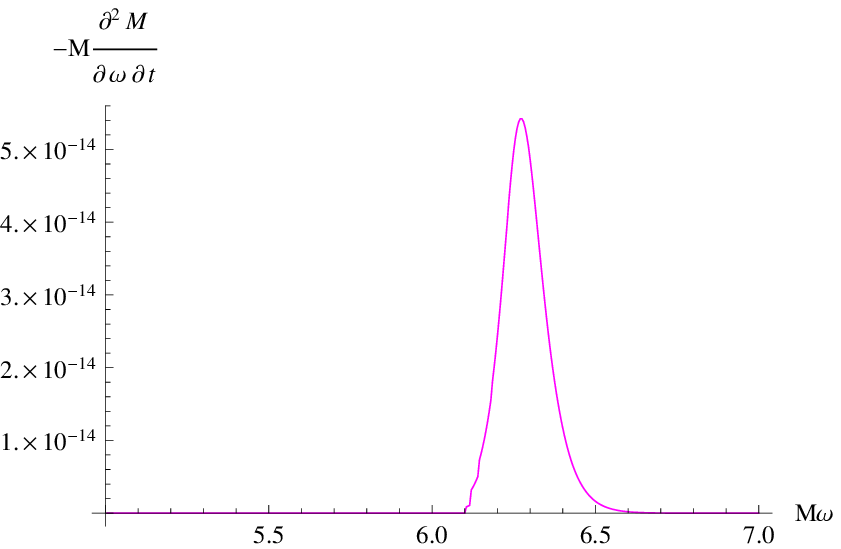}}
\caption{Energy-emission rate for the Reissner-Nordstr\"om black hole ($r_-\approx0$) of charged scalar particles and antiparticles for $M\mu=1$  $eQ=\pm1$(left panel) and $M\mu=5$ $eQ=\pm10$ (right panel).
}\label{fig:Erate}
\end{figure*}

\begin{figure*}
\resizebox{\linewidth}{!}{\includegraphics*{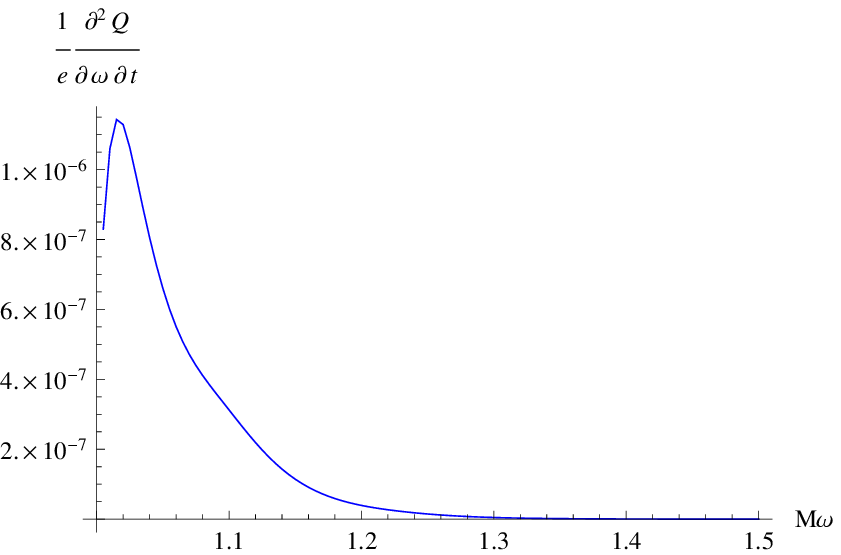}\includegraphics*{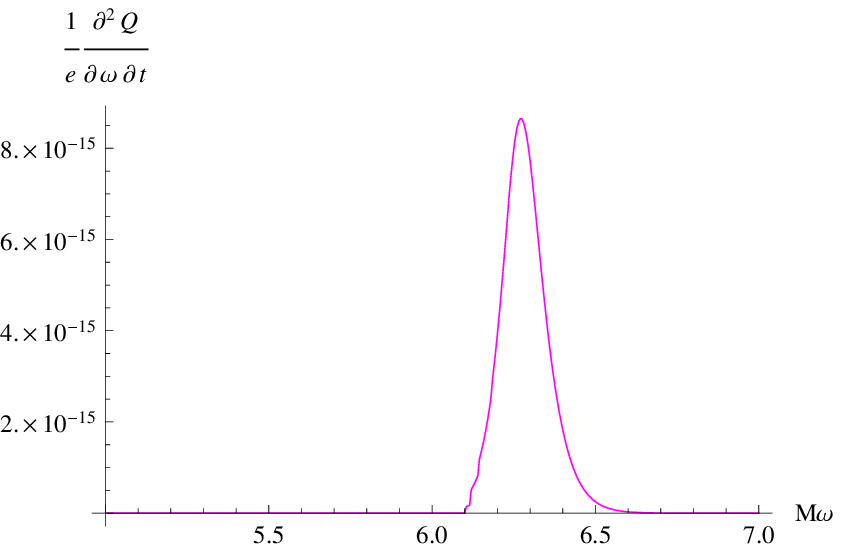}}
\caption{Charge-emission rate for the Reissner-Nordstr\"om black hole ($r_-\approx0$) of charged scalar particles and antiparticles for $M\mu=1$  $eQ=\pm1$(left panel) and $M\mu=5$ $eQ=\pm10$ (right panel).
}\label{fig:Crate}
\end{figure*}

\begin{figure*}
\resizebox{\linewidth}{!}{\includegraphics*{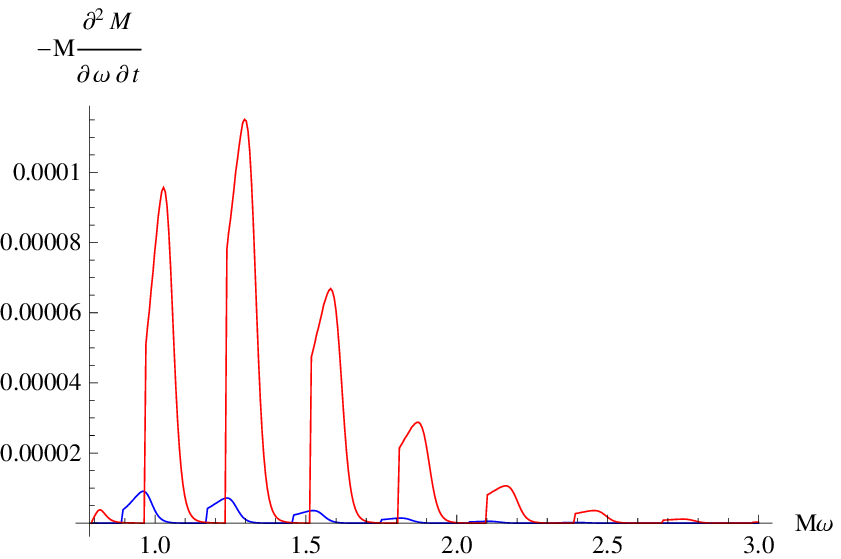}\includegraphics*{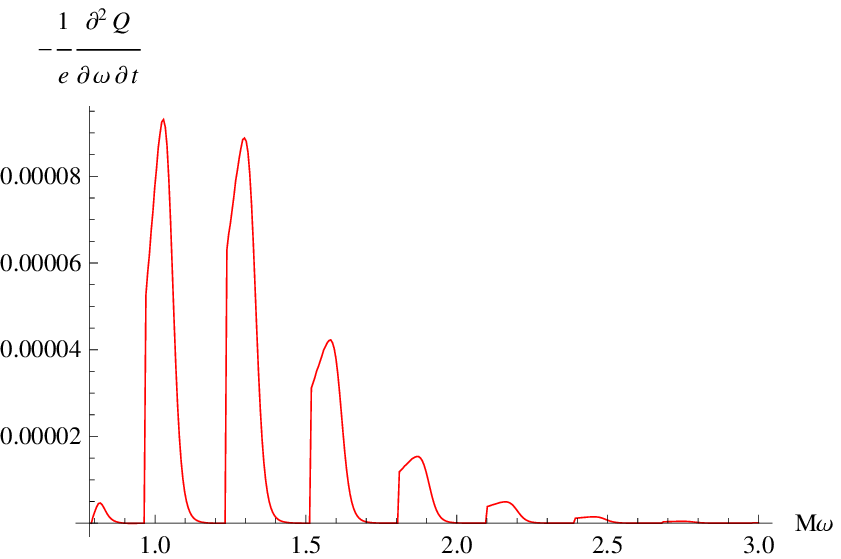}\includegraphics*{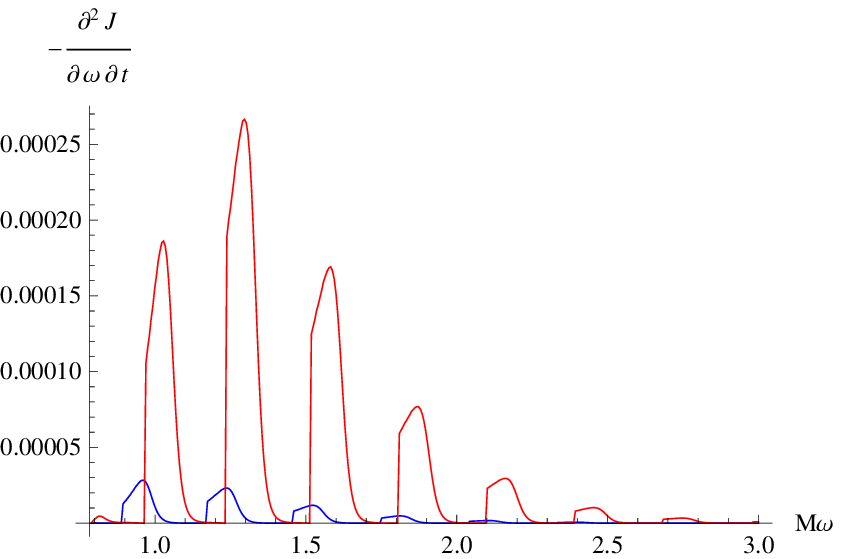}}
\caption{Energy-, momentum-, and charge-emission rates for the Kerr-Newman black hole ($r_-/r_+=a^2/r_+^2=0.49$, $J/M^2=a/M\approx0.94$) of the scalar particles and antiparticles ($\mu M=1$): $eQ=1$ (red, upper) and $e=0$ (blue, lower).
}\label{fig:rot}
\end{figure*}

\begin{table}
\caption{Energy-emission and charge-emission rates for the Reissner-Nordstr\"om black hole ($r_-\approx0$) of charged scalar particles and antiparticles. For the comparison, energy-emission rate due to a massless uncharged scalar field is $M^2\frac{dM}{dt}=1.4859\times10^{-4}$.}\label{tabl:rates}
\begin{tabular}{|r|l|l|l|l|}
\multicolumn{1}{l}{} & \multicolumn{2}{c|}{$M\mu=1$} & \multicolumn{2}{c}{$M\mu=5$}\\
\hline
$|eQ|$&$M^2\frac{dM}{dt}$&$\frac{M}{e}\frac{dQ}{dt}$&$M^2\frac{dM}{dt}$&$\frac{M}{e}\frac{dQ}{dt}$\\
\hline
$0$&$2.910\times10^{-12}$&$0$&$6.951\times10^{-54}$&$0$\\
$1$&$8.937\times10^{-8}$&$8.433\times10^{-8}$&$8.078\times10^{-49}$&$1.602\times10^{-49}$\\
$5$&$7.205\times10^{-3}$&$2.768\times10^{-3}$&$8.519\times10^{-28}$&$1.687\times10^{-28}$\\
$10$&$0.0689$&$0.0136$&$8.830\times10^{-15}$&$1.405\times10^{-15}$\\
\hline
\end{tabular}
\end{table}

\begin{table}
\caption{Energy-, momentum-, and charge-emission rates for the Kerr-Newman black hole ($r_-/r_+ \thickapprox a^2/r_+^2=0.49$, $J/M^2=a/M\approx0.94$) of the scalar particles and antiparticles ($\mu M=1$).}\label{tabl:ratesrot}
\begin{tabular}{|r|l|l|l|}
\hline
$|eQ|$&$M^2\frac{dM}{dt}$&$M\frac{dJ}{dt}$&$\frac{M}{e}\frac{dQ}{dt}$\\
\hline
$0$&$2.038\times10^{-6}$&$6.574\times10^{-6}$&$0$\\
$1$&$3.106\times10^{-5}$&$7.221\times10^{-5}$&$2.358\times10^{-5}$\\
\hline
\end{tabular}
\end{table}

Not all positive energy particles can escape the black hole, as part of them is reflected from the potential barrier back to the black hole. Consequently, the emission rates, given by the formulas above, depend on the grey-body factors $\left| {\cal A}_{\ell, m} \right|^2$  which give the fraction of particles penetrating the barrier.

Massive fields have two extra features: modes with $\omega < \mu$ naturally do not contribute into the radiation process. Within the existing semiclassical approach one cannot trust the results in the regime $\mu M \ll 1$, as the backreaction of the quantum fields may be not small.

In order to distinguish the effect of the charge itself, first, we shall consider charged nonrotating (Reissner-Nordstr\"om) black holes. Computations of the energy- and charge-emission rates are shown on Figs. 1 and 2. There one can see that the ``mass term'' $\mu M$ strongly suppresses the radiation of energy and charge. Once we assume that the initial black hole charge cannot normally be much larger then the accreting limit $\mu M/e$, the discharge occurs slowly, as can be noticed from small emission rates on Fig. \ref{fig:Crate} and Table \ref{tabl:rates}.  The latter statement is not in any kind of contradiction with the classical papers \cite{Carter,Gibbons:1975kk,Page:1977um}, claiming that the black hole discharges quickly, because an initial black hole charge considered there was supposed to be much larger then its accreting limit $Q \gg \mu M/e$. Here we do not suppose an intense external ``stuffing'' of an electric charge into the black hole. Alternatively, if someone, anyway, implies a large initial charge $Q$, our results here simply refer to later stages of evaporation, when the initial intense discharge has damped. Despite relatively small values of the black hole charge, this later stage of the discharge is not insignificant for electromagnetic processes in the black hole's vicinity, as the coupling $e Q$ does not need to be not small.

From Fig. \ref{fig:rot} and Table \ref{tabl:ratesrot}, one can see that high rotation can enhance the emission of mass, angular momentum, and charge up to a few orders. Therefore, the four-dimensional black hole should first lose the most part of its angular momentum, and then, discharge.

\section{$Q/M$ evolution due to evaporation}

Even though we are using here a quasistationary model of the black hole evaporation, let us try to speculate on what may happen with the ratio $Q/M$ during this relatively early period of Hawking radiation, that is, when the semiclassical regime is still valid.

\begin{eqnarray}
M\frac{d}{dt}\frac{Q}{M}&=&\frac{dQ}{dt}-\frac{Q}{M}\frac{dM}{dt}=\nonumber\\
&=&\frac{1}{MQ}\left(eQ\frac{M}{e}\frac{dQ}{dt}-\frac{Q^2}{M^2}M^2\frac{dM}{dt}\right),\label{QoM}
\end{eqnarray}
where $\frac{dQ}{dt}$ and $\frac{dM}{dt}$ are negative due to the Hawking radiation.

If $Q/M$ is large enough the second term in (\ref{QoM}) is dominant, and the absolute value of $Q/M$ increases due to the radiation. Below some value of $Q/M$ the first term is dominant and $Q/M$ decreases until the black hole discharges. However, as discussed in the introduction, we consider values of $Q/M \sim \mu/e \sim 10^{-21}$. 
In this regime, from the table~\ref{tabl:rates} we deduce that the behaviour is more complicated: the second term appears to be dominant for $eQ=1$ and $\mu M=5$ because $M^2\frac{dM}{dt}\sim10^{-3}$ due to all the Standard Model particles radiation, and $\frac{M}{e}\frac{dQ}{dt}\sim10^{-49}$ which is less than the factor $\frac{Q^2}{M^2}\sim10^{-42}$. In practice this means that the probability to lose even one electron (positron) for such a black hole is negligible. As $Q$ remains constant and $M$ decreases due to the Hawking radiation the absolute value of $Q/M$ increases. However for $\mu M=1$ we observe that the first term in (\ref{QoM}) is again dominant and the black hole discharges. Therefore, we conclude that the primordial black hole can possess small charges (of order of tenths of the elementary charge $e$) until $eQ<\mu M$ (see the charge-emission rate for $eQ\geq\mu M$ in table~\ref{tabl:rates}). After the black hole mass $M$ becomes smaller than $eQ/\mu$, charged particles are more probable to be emitted until the charge-to-mass ratio becomes again smaller and of order $Q/M\lesssim10^{-21}$.

Since $\mu M\geq1$, the ``excess'' of charge $Q > \mu M/e$ is likely to be radiated away sooner than the ration $Q/M$ will reach the threshold of the Schwinger effect $e Q \sim (\mu M)^2$. Thus, the Schwinger effect should not be significant while the system is within the semiclassical regime.

Let us estimate a characteristic time $\Delta t$ elapsed between two consequent emissions of charged particles. Suppose that at some moment an electron (positron) was emitted by a black hole with an initial charge $Q_0=Ne$, $N\sim10^2\gg1$, then the black hole charge decreased by $e$, i.e., $Q_1=(N-1)e$ and the second term became dominant. In order to ``restore'' the ratio between $eQ$ and $\mu M$ (as shown above), the black hole must be evaporating for some time (without emitting any charge) until its mass $M_0$ falls down to $M_1=\frac{N-1}{N}M_0$, due to the radiation of all the particles. Since $N\gg1$, $Q_1/Q_0\approx1$ and, once the mass $M_1$ is reached, the probability of the next charged particle's emission should be approximately the same as before the emission of the first charged particle.

Because of all the fields' radiation, the black hole mass changes as
$$1-\frac{M^3}{M_0^3}=\frac{t}{\tau},$$
where $\tau$ is the characteristic lifetime of the black hole.

For $M=M_1$ we obtain
$$\Delta t= \tau\left(1-\frac{M_1^3}{M_0^3}\right)=\tau\left(1-\frac{(N-1)^3}{N^3}\right)\approx\frac{3\tau}{N}.$$

If we take, say, $10^{10}$ years, as a typical lifetime estimation for primordial black holes \cite{Page:1977um}, even for the ``high'' initial charge $Q \sim 10^2 e$ the emission of electron (positron) is a very rare phenomenon, which will occur approximately in $10^{8}$ years. Thus, a primordial black hole of mass $M \sim 10^{15} - 10^{14} kg.$ has a quasiconstant electric charge during a very long period, while its mass is intensively evaporating. This ``fragmentary'' picture of discharging is certainly valid only within semiclassical regime and should be corrected at the latest stage of evaporation. Moreover, once we are concerned about these particular, ``fragmentary'' events of the charge's emission, a rigorous quantum filed theory (allowing one to compute ``probabilities of particular events'' rather than ``average probabilities'') in the black hole background should come in place of the conventual semiclassical treatment.

\section{Conclusions}

In this paper we have filled the existing gap in understanding the process of the discharge of the Kerr-Newman black hole owing to the Hawking radiation of charged particles. Unlike previous calculations of the intensity of Hawking radiation, which were made either for $\mu M \ll 1$ (when one cannot trust to semiclassical approximation) or for $\mu M \gg 1$ (when the evaporation of massive fields is greatly suppressed), we analyzed the case $\mu M \gtrsim 1$. We have shown that in this regime, the discharge of a black hole occurs in a kind of ``fragmentary'' way with a very large interval between two consequent emissions of charged particles. Rotation greatly enhances the emissions rates.

An accurate and full, semiclassical picture of the Kerr-Newman black hole evaporation can be obtained after a similar analysis of the Hawking evaporation of charged fermions \cite{progress}, which is the subject of our future research.

\section*{Acknowledgments}
R. A. K. acknowledges the support of his visit to Universidade Federal do ABC by FAPESP.
A.~Z. was supported by the Alexander von Humboldt Foundation, Germany and Coordena\c{c}\~ao de Aperfei\c{c}oamento de Pessoal de N\'ivel Superior (CAPES), Brazil.

\end{document}